\documentclass[fleqn,12pt,twoside]{article}
\usepackage{espcrc1}

\usepackage{graphicx}

\usepackage[figuresright]{rotating}

\newcommand{\AmS}{{\protect\the\textfont2
  A\kern-.1667em\lower.5ex\hbox{M}\kern-.125emS}}

\newcommand{\be}{\begin{equation}}
\newcommand{\ee}{\end{equation}}
\newcommand{\bea}{\begin{eqnarray}}
\newcommand{\eea}{\end{eqnarray}}
\newcommand{\bt}{\begin{tabular}}
\newcommand{\et}{\end{tabular}}

\newcommand{\rt}{\rightarrow}

\newcommand{\etal}{{\it et al.}}

\title{The Nuclear Reactions in Standard BBN}

\author{Pasquale D. Serpico
\address[MCSD]{Max-Planck-Institut f\"ur
Physik (Werner-Heisenberg-Institut),\\ F\"ohringer Ring 6, 80805
M\"unchen, Germany}\\ \texttt{serpico@mppmu.mpg.de}}

\begin{document}

\maketitle

\begin{abstract}
Nowadays, the Cosmic Microwave Background (CMB) anisotropy
studies accurately determine the baryon fraction $\omega_b$,
showing an overall and striking agreement with previous
determinations of $\omega_b$ obtained from Big Bang
Nucleosynthesis (BBN). However, a deeper comparison of BBN
predictions with the determinations of the primordial light
nuclide abundances shows some tension, motivating an effort to
further improve the accuracy of theoretical predictions, as well
as to better evaluate systematics in both observations and nuclear
reactions measurements. We present some results of an important
step towards an increasing precision of BBN predictions, namely an
updated and critical review of the nuclear network, and a new
protocol to perform the nuclear data regression.
\end{abstract}

\section{Introduction}
In the framework of the ``Cosmological Concordance Model'',
BBN probes the earliest times, but once $\omega_b$ is fixed, as
now allowed by accurate CMB anisotropies
analysis~\cite{Spergel:2003cb}, it is completely ruled by Standard
Physics, at least in its minimal formulation, thus providing
useful insights on a wide range of astrophysical and cosmological
issues.

As well known, the nuclide abundance predictions are mainly
affected by the nuclear reaction uncertainties. The suggested
range of $\omega_b$ and the low-energy nuclear data taken in the
last decade justify a revision of the BBN network reliability,
with respect to the one performed in the seminal paper by Smith
\etal~\cite{Smith:1992yy}: here we summarize our techniques and
some results. For a detailed discussion of the issues presented
here and of other refinements (like a better treatment of the
neutrino sector and of plasma and QED effects) see~\cite{bbnnuclI}.

\section{From the data to the rates}

The reaction rates are obtained as thermal averages of the
relevant astrophysical factors $S(E)$ \footnote{In what follows
for simplicity we will refer to the S-factor, clearly defined only
for reactions induced by charged particles. Similar relations also
hold for neutron induced reactions.}, so the first and most
critical step is how to combine the data $\{i_k\}$ of several
experiments $\{k\}$, each one affected in general by statistical,
$\sigma_{i_k}$, and normalization errors, $\epsilon_k$, in a
meta-analysis where both the magnitude and the energy behavior of
the $S(E)$ have to be deduced. Usually, different experiments
disagree within the quoted errors, suggesting some systematic
discrepancy, the bulk of which can be attributed to different
normalizations. There is obviously no unique and unambiguous way
to deal with such discrepancies and several methods appeared in
the literature. In a recent one described in~\cite{Cyburt:2004cq},
both the errors are included in the covariance matrix entering the
expression of the $\chi^2$ function to be minimized. This approach
clearly takes into account the correlations between data of the
same experiment, but as addressed in~\cite{D'Agostini:1993uj}, a
bias is thus introduced, leading to a systematic underestimate of
the fitted functions, that typically pass below the majority of
the data, and possibly affecting the nuclide predictions and
errors estimates.

\begin{figure}[htb]
\begin{minipage}[t]{70mm}
\includegraphics[width=6.94cm]{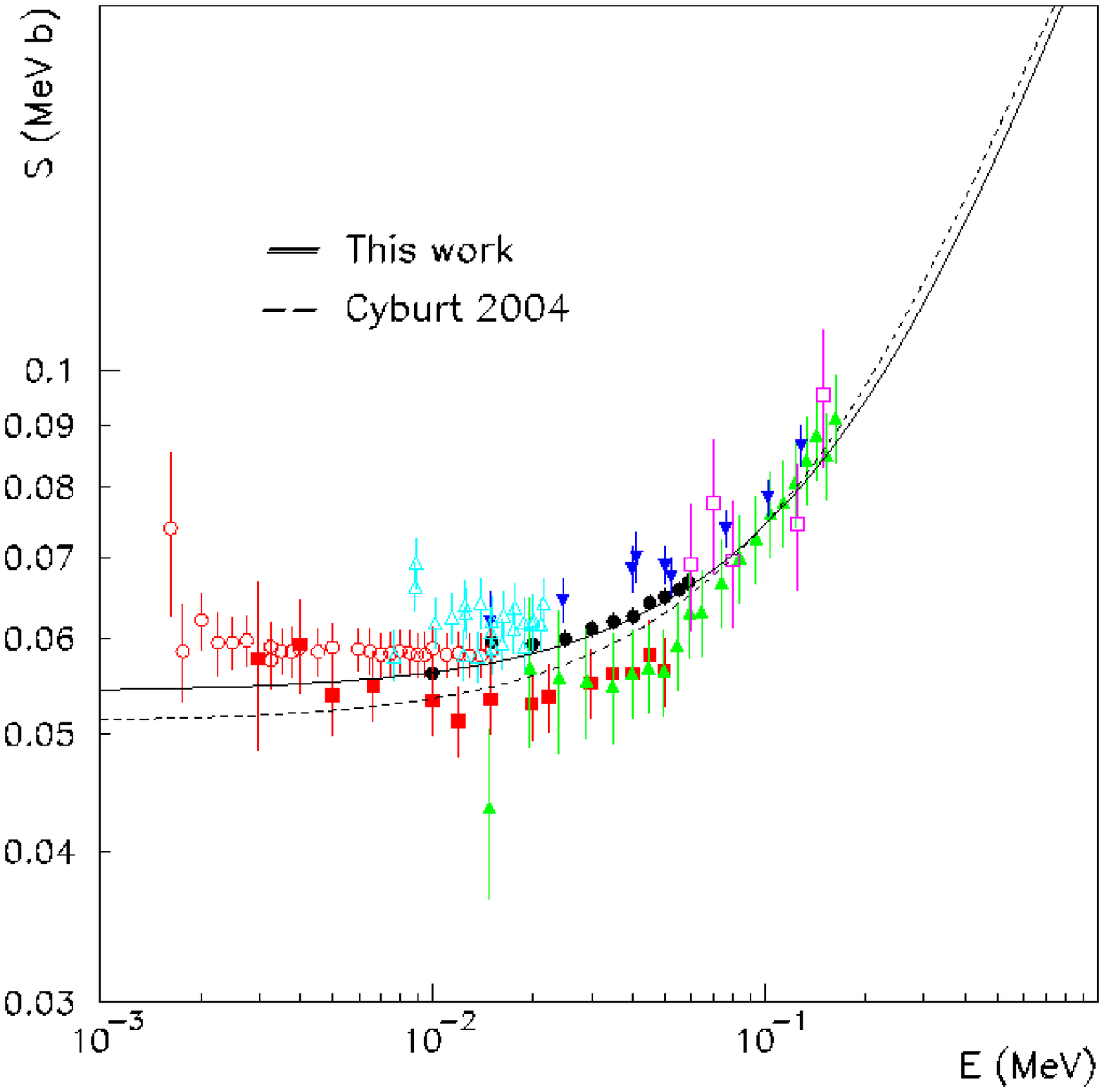}
\caption{Comparison between our regression method (solid line) and
the one in ref~\cite{Cyburt:2004cq}(dashed line)
 for the $^2$H(d,p)$^3$H reaction.} \label{fig:ddp}
\end{minipage}
\hspace{\fill}
\begin{minipage}[t]{70mm}
\includegraphics[width=6.88cm]{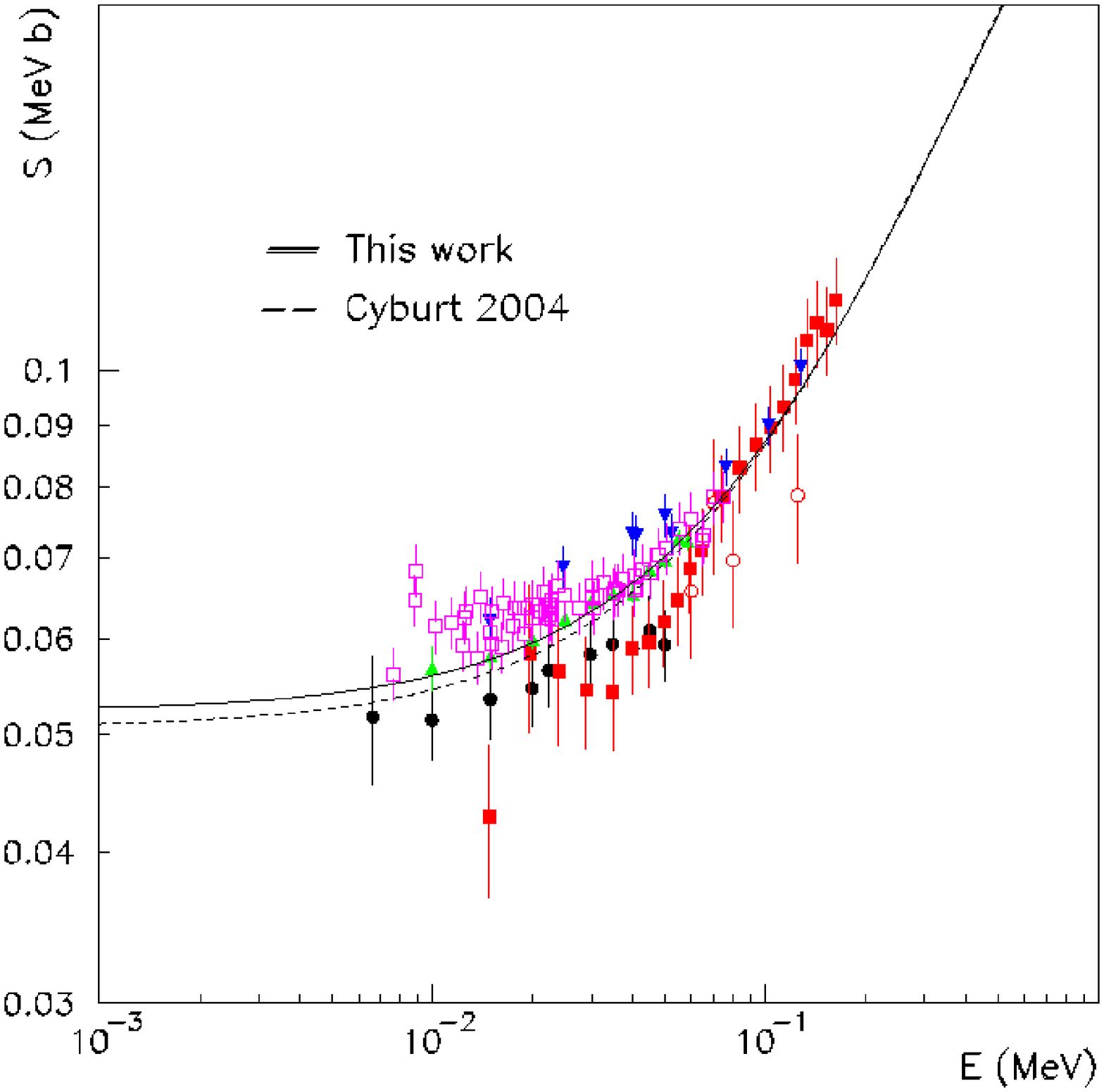}
\caption{As in Fig.\ref{fig:ddp}, but for the $^2$H(d,n)$^3$He
process. Different symbols are used to denote different data
sets.} \label{fig:ddn}
\end{minipage}
\end{figure}

In our method, that generalizes the
approach suggested in~\cite{D'Agostini:1993uj}, the $\chi^2$ is calculated as
\begin{equation}
\chi^2(a_l,\omega_k)=\chi_{stat}^2+\chi_{norm}^2\equiv \sum_{i_k}
\frac{(S_{th}(E_{i_k},a_l)-S_{i_k}\omega_k)^2}{\omega_k^2\sigma_{i_k}^2}+
\sum_{i_k}\frac{(\omega_k-1)^2}{\epsilon_{i_k}^2},
\end{equation}
where $S_{i_k}$ is the data point at the energy $E_{i_k}$ and
$S_{th}(E)$ is the theoretical value of the astrophysical factor,
depending on some parameters $a_l$ to be determined together with
the renormalizing factors $\omega_k$ by standard minimization
procedures. The covariance matrix was built by considering the
statistical errors only\footnote{Or conservatively the total ones,
if $\epsilon_k$ and $\sigma_{i_k}$ were unavailable in a separate
form.}, while different renormalization factors were allowed for
each data set; the introduction of $\chi_{norm}^2$ disfavors
renormalizations greater than the estimated $\epsilon_k$. The
best-fit curves thus produced pass \emph{through} the data, closer
to the best determinations, as one would expect from an unbiased
estimator (see Figs.~\ref{fig:ddp},\ref{fig:ddn}). In such an
approach, the bulk of the systematic uncertainty has been taken
into account and the residual discrepancies can be considered as
due to some unidentified/underestimated source of error in one or
several experiments. Then, we simply inflated the calculated error
by a scale factor $\sqrt{\chi_\nu^2}$, as prescribed by the
Particle Data Book~\cite{Eidelman}. Using this prescription, 
the overall normalization uncertainty cannot be clearly worse 
than the one determined by the most accurate experiment: this 
error was quadratically added to the previous one. Finally, a 
typical renormalization factor $\varepsilon$ was calculated, 
to give an idea of the current disagreement on the scale of $S(E)$ 
among several experiments
\begin{equation}
\varepsilon^2\equiv \frac{\sum_k w_k(\omega_k-1)^2}{\sum_k w_k}
\end{equation}
where $w_k=(\chi_k^2/N_k)^{-1}$, $N_k$ is the number of data and
$\chi_k^2$ the contribution to the $\chi^2$ of the k-th data set,
thus assigning a larger weight to the experiments closer to the
fitted $S$.

The rate $R$ was obtained by numerical integration of $S(E)$
convolved with the Boltzmann/Gamow Kernel $K(E,T)\sim
e^{-E/T}e^{-\sqrt{E_G/E}}$
\begin{equation}
R =\int_{0}^{\infty}dE\: K(E,T)\:S(E,\hat{a})
\end{equation}
and its error $\delta R$ trough the standard error propagation as
\begin{equation}
\delta R^2 =\int_{0}^{\infty} dE^{'} K(E^{'},T) \int_{0}^{\infty}
 dE K(E,T) \sum_{l,m}\frac{\partial S(E^{'},a)}{\partial a_l}\bigg|_{\hat{a}}
\frac{\partial S(E,a)}{\partial a_m}\bigg|_{\hat{a}} cov(a_l,a_m),
\end{equation}
thus fully including the correlations among the fitted parameters.
The uncertainties on the nuclides yields were finally obtained
with a generalization of the linear propagation method described
in~\cite{bbnnuclI,Cuoco:2003cu,Serpico:2004xy}.

\section{From the rates to the predictions}

\begin{figure}[htb]
\begin{minipage}[t]{72mm}
\includegraphics[width=7.05cm]{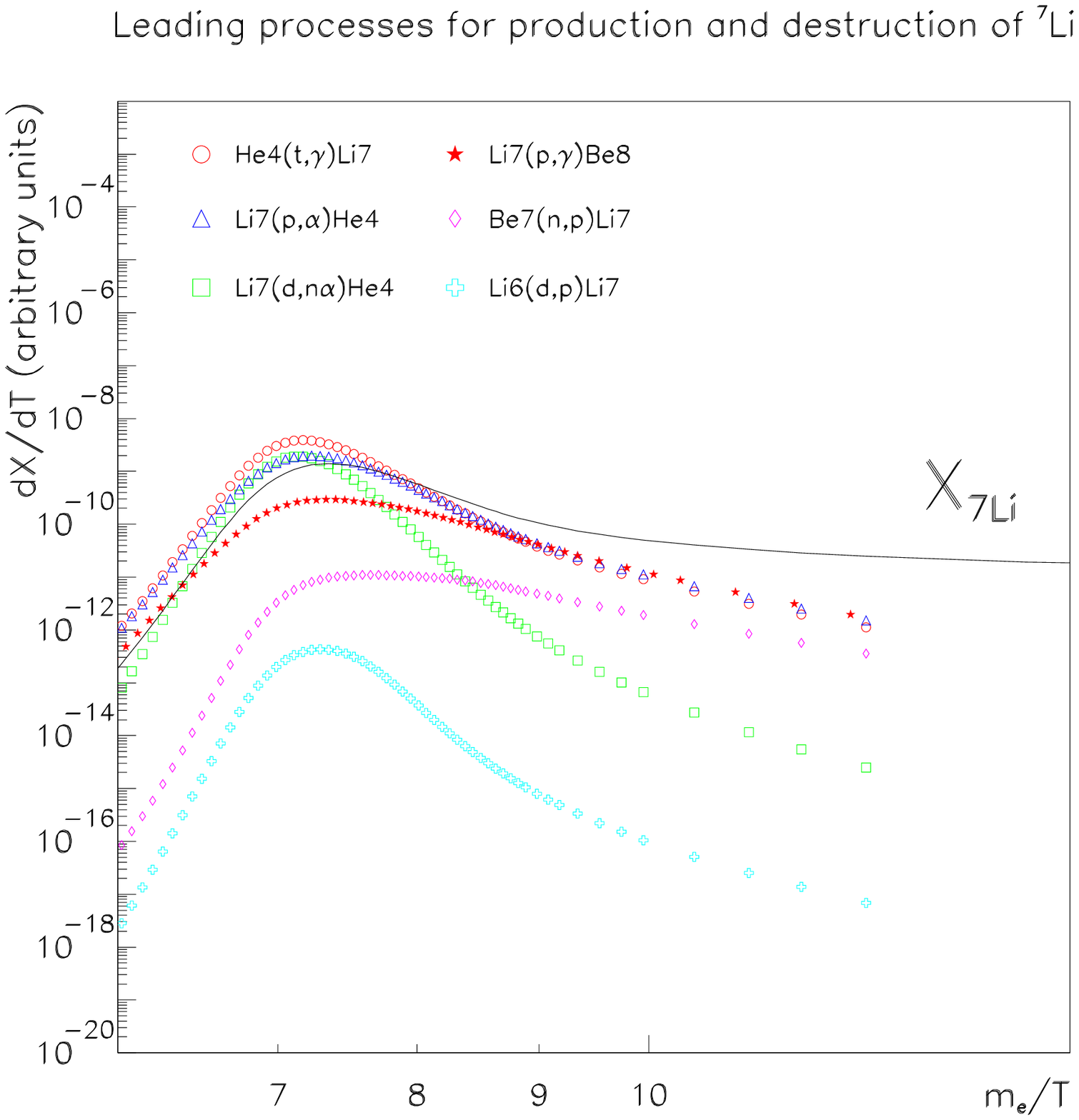}
\caption{The r.h.s. of the Boltzmann equations for the reactions
relevant for the synthesis and destruction of $^7$Li.}
\label{fig:li7}
\end{minipage}
\hspace{\fill}
\begin{minipage}[t]{72mm}
\includegraphics[width=7.18cm]{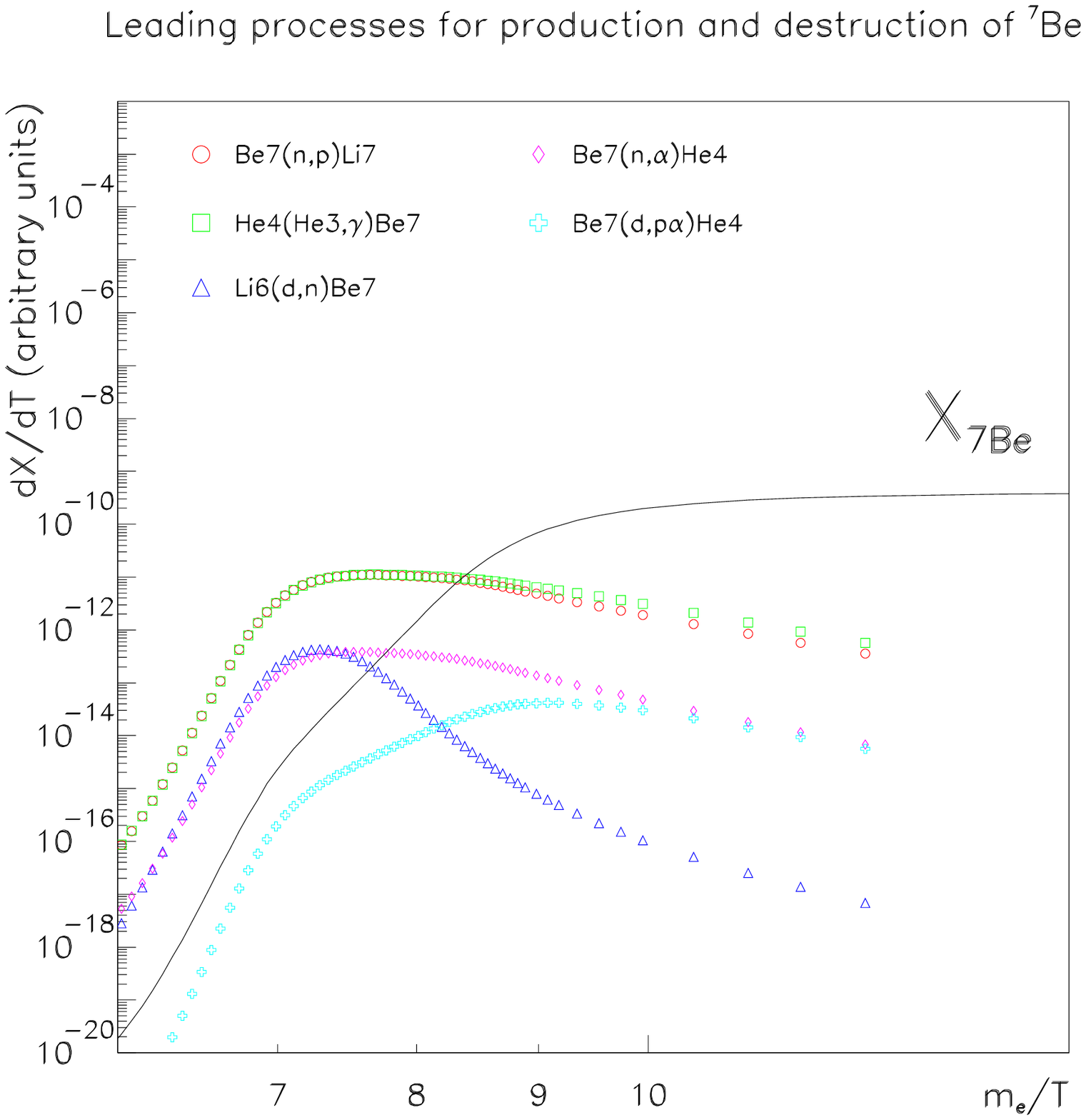}
\caption{As in Fig.\ref{fig:li7}, but for $^7$Be. Like in
Fig.\ref{fig:li7}, the nuclide abundance evolution is also shown.}
\label{fig:be7}
\end{minipage}
\end{figure}

In Fig.~\ref{fig:li7} (\ref{fig:be7}) we plot the net contribution
to the right-hand side of the leading and main sub-leading
reactions to the synthesis and destruction of $^7$Li ($^7$Be),
obtained for the typical value $\omega_b=0.023$. It is easily seen
that the first stage of $^7$Li production around $T\simeq 70$ keV
mainly proceeds through the $^4$He(t,$\gamma$)$^7$Li reaction, but
$^7$Li is soon burned via $^7$Li(p,$\alpha$)$^4$He reaching a
low plateau value. The final $^7$Li content is thus mainly given
by the late $^7$Be EC decay: this isotope, synthesized via the key
route $^4$He($^3$He,$\gamma$)$^7$Be, is less easily destroyed
through the (experimentally poorly known) channels
$^7$Be(n,$\alpha$)$^4$He and $^7$Be(d,p)2$^4$He \footnote{The
latter has been measured at Louvain la Neuve, see the contribution of C.
Angulo to these proceedings.}, or the two steps process
$^7$Be(n,p)$^7$Li + $^7$Li(p,$\alpha$)$^4$He, now determined with
good accuracy. The former two reactions are indeed responsible for
the bulk of the $^7$Li error budget, at least if a conservative
uncertainty of one order of magnitude is assumed.

We also found that sub-leading processes as $^7$Li+p$\rt$
$^8$Be$^{*}\rt\gamma$+2 $^4$He, often neglected in BBN studies,
for which new measurements exists, are truly marginal, even if
their little contribution to the final error is similar to the
much widely treated $^4$He(t,$\gamma$)$^7$Li. The same is true
e.g. for the other overlooked reactions $^6$Li(d,p)$^7$Li and
$^6$Li(d,n)$^7$Be.

We checked that only a handful of reactions dominate the error
budget, and, apart for useful measurements of poorly known cross
sections, a determination of \emph{both the magnitude and the
shape} of the $^2$H(d,n)$^3$He,$^2$H(d,p)$^3$H,
$^4$He($^3$He,$\gamma$)$^7$Be reactions at the 1\% accuracy level
over \emph{all} the interval of interest for the BBN (say, up to
$\sim$ 2 MeV) could significantly improve the reliability of the
predictions of both $^2$H and $^7$Li. In these cases, indeed, the
systematics coming from several experiments dominate the
uncertainty, and more reliable data are needed, since even very
detailed regression methods may fail in these cases. On the other
hand, the BBN theory would surely benefit of refined studies of
the $^4$He and $^7$Li observational systematics, as well as from
an increase in the statistics of the $^2$H/H determinations in the
high-z damped Ly-$\alpha$ absorbtion systems.

We also confirm that the $^4$He error is dominated by that on the
neutron lifetime, while the new value of $G_N$ quoted
in~\cite{Eidelman} makes its uncertainty of no relevance for the
BBN.

In summary, we presented some highlights on a new method of data
regression and a reanalysis and update of the BBN nuclear network.
Some differences with the results currently quoted in the
literature were discussed, but typically a reassuring agreement
with the usual results has been found, confirming the robustness
of the BBN predictions.

Obviously, the new compilation produced will also turn to be
useful to deepen our insights on several non standard BBN scenarios.


\begin{thebibliography}{10}

\bibitem{Spergel:2003cb}
D.~N.~Spergel {\it et al.},
Astrophys.\ J.\ Suppl.\  {\bf 148} (2003) 175
[arXiv:astro-ph/0302209].


\bibitem{Smith:1992yy}
M.~S.~Smith, L.~H.~Kawano and R.~A.~Malaney,
Astrophys.\ J.\ Suppl.\  {\bf 85} (1993) 219.

\bibitem{bbnnuclI}
P.~D.~Serpico \etal,
arXiv:astro-ph/0408076.


\bibitem{Cyburt:2004cq}
R.~H.~Cyburt,
Phys.\ Rev.\ D {\bf 70}, 023505 (2004)
[arXiv:astro-ph/0401091].

\bibitem{D'Agostini:1993uj}
G.~D'Agostini,
Nucl.\ Instrum.\ Meth.\ A {\bf 346} (1994) 306.

\bibitem{Eidelman}
S. Eidelman \etal, Phys. Lett. {\bf B 592} (2004) 1.

\bibitem{Cuoco:2003cu}
A.~Cuoco \etal,
arXiv:astro-ph/0307213, to appear in Int.Journ.Mod.Phys.A.

\bibitem{Serpico:2004xy}
P.~D.~Serpico,
arXiv:astro-ph/0401072.

\end{thebibliography}
\end{document}